\documentclass[aps,prd,preprint,showpacs,superscriptaddress]{revtex4-1}

\usepackage{epsfig}
\usepackage{graphicx}
\usepackage{dcolumn}
\usepackage{bm}
\usepackage{multirow}
\usepackage{array}
\usepackage{slashed}
\usepackage{amsmath}
\usepackage{amssymb}
\usepackage[titletoc,page]{appendix}
\usepackage{color}
\usepackage{hyperref}

\begin{document}

\title{Spectra and rates of bremsstrahlung neutrino emission in stars} 

\author{Gang Guo}
\affiliation{Center for Nuclear Astrophysics, Department of Physics and Astronomy, Shanghai Jiao Tong University, Shanghai 200240, China}
\author{Yong-Zhong Qian}
\email{qian@physics.umn.edu} 
\affiliation{School of Physics and Astronomy, University of Minnesota, Minneapolis, MN 55455, USA}
\affiliation{Center for Nuclear Astrophysics, Department of Physics and Astronomy, Shanghai Jiao Tong University, Shanghai 200240, China}

\date{\today}

\begin{abstract}
We calculate the energy-differential rate for neutrino emission from 
electron-nucleus bremsstrahlung in stellar interiors taking into account the effects 
of electron screening and ionic correlations. We compare the energy-differential and 
the net rates, as well as the average $\bar\nu_e$ and $\bar\nu_x\ (x=\mu,\tau)$ 
energies, for this process with those for $e^\pm$ pair annihilation, plasmon decay, 
and photo-neutrino emission over a wide range of temperature and density. We also 
compare our updated energy loss rates for the above thermal neutrino emission 
processes with the fitting formulas widely used in stellar evolution models and 
determine the temperature and density domain in which each process dominates. 
We discuss the implications of our results for detection 
of $\bar\nu_e$ from massive stars during their pre-supernova evolution and find 
that pair annihilation makes the predominant contribution to the signal 
from the thermal emission processes.
\end{abstract}   


\maketitle

\section{Introduction} \label{sec:intro}
Stars are profuse sources of neutrinos. A prominent example is the solar 
neutrinos produced by weak nuclear reactions including electron capture
and $\beta$ decay. For stars like the sun and those of higher masses, as 
temperature and density increase during later stages of their 
evolution, $\nu\bar\nu$ pair production by $e^\pm$ pair annihilation, 
plasmon decay, photo-neutrino emission, and electron-nucleus 
bremsstrahlung becomes more and more important. Indeed, for those 
stars that can ignite core carbon (C) burning, the energy loss subsequent 
to C ignition is dominated by the above so-called thermal neutrino 
emission processes. For stars of $\gtrsim 8\,M_\odot$ ($M_\odot$ being
the mass of the sun), neutrinos 
not only drive their evolution by cooling their interiors but also play 
dynamic roles in their core collapse and the ensuing supernova explosion.

The thermal neutrino emission processes in stellar interiors have been 
studied extensively \cite{BPS1967,Dicus72,Schinder87,%
Itoh85,Itoh86,Itoh89,Itoh92,Itoh93,Itoh94,Itoh96a,Itoh96b}. As neutrinos 
free-stream out of massive stars during their pre-supernova evolution, the
pertinent quantity for stellar evolution is the energy loss rate of each process.
Practically, fitting formulas for these rates given by Ref.~\cite{Itoh96a}
have been widely used in stellar evolution models. As thermal neutrino 
emission depends on temperature and density and evolves as stars age, 
these neutrinos would constitute a unique probe of the conditions in stellar 
interiors, thereby providing a potential test of stellar evolution models 
\cite{Odrzywolek10,Kato15,JUNO,Yoshida:2016}. Even if there might not be sufficient 
statistics to probe the details of stellar evolution, unambiguous detection 
of pre-supernova neutrinos from a nearby massive star would at least 
provide advance warning for the subsequent supernova explosion 
\cite{Odrzywolek04,KamLAND15,Yoshida:2016}. For the above purposes, it is
important to calculate the detailed spectra of the thermal neutrino emission
processes. In addition, neutrino signals from massive stars during their
pre-supernova evolution are affected by flavor transformation through the
Mikheyev-Smirnov-Wolfenstein (MSW) mechanism \cite{MSW1,MSW2}. 
A careful analysis of the MSW effect on these neutrino signals also requires 
knowledge of the neutrino spectra.
 
The neutrino spectra for the thermal emission processes can be obtained from 
the corresponding energy-differential rates. Previous works
\cite{Odrzywolek04,Ratkovi03,Dutta03,Misiaszek05,Odrzywolek07,Kato15,Patton15} have
studied the neutrino spectra for $e^\pm$ pair annihilation, plasmon decay, 
and photo-neutrino emission. The spectra for electron-nucleus bremsstrahlung 
have not received as much attention. In particular, we are not aware of 
a detailed comparison of the spectra for this and other thermal emission 
processes. In this paper we focus on neutrino emission from electron-nucleus 
bremsstrahlung in massive stars and its importance relative to other processes.
Following a detailed comparison of the energy-differential rates of all the 
thermal neutrino emission processes during the pre-supernova evolution of a
massive star, we find that $e^\pm$ pair annihilation makes the predominant 
contribution to the $\bar\nu_e$ signal from these processes for detection 
through capture on protons.

We present a detailed derivation of the energy-differential rate 
for neutrino emission from electron-nucleus bremsstrahlung in Sec.~\ref{sec:brem}.
We compare the energy-differential and the net rates, as well as the average 
$\bar\nu_e$ and $\bar\nu_x\ (x=\mu,\tau)$ energies, for this and other thermal 
neutrino emission processes over a wide range of temperature and density in 
Sec.~\ref{sec:spec}. We also compare our updated energy loss rates for 
individual thermal neutrino emission processes with the fitting 
formulas of Ref.~\cite{Itoh96a} and determine the temperature and density 
domain in which each process dominates in Sec.~\ref{sec:loss}. We discuss 
the implications of our results for detection of $\bar\nu_e$ from massive stars 
during their pre-supernova evolution and give conclusions in Sec.~\ref{sec:disc}.

\section{Energy-differential rates for bremsstrahlung neutrino emission}
\label{sec:brem}
\begin{figure}[htbp]
\centering
\includegraphics[width=8.cm]{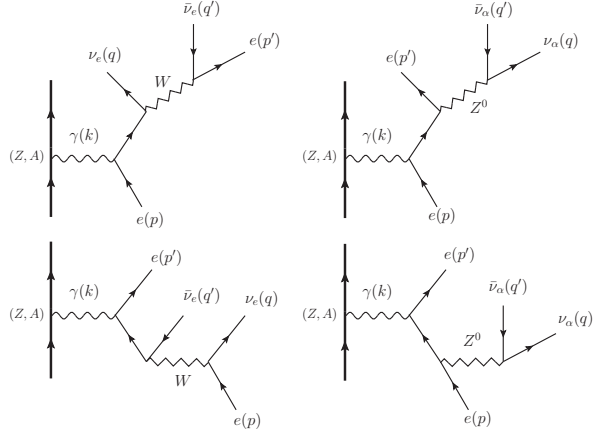}
\caption{Leading-order Feynman diagrams for neutrino emission from electron-nucleus 
bremsstrahlung.}
\label{fig:brem}
\end{figure}

Neutrino emission from electron-nucleus bremsstrahlung is denoted by
\begin{equation}
(Z, A) + e^- \to (Z, A) + e^- + \nu_\alpha + \bar\nu_\alpha,
\end{equation}
where $(Z,A)$ represents a nucleus of proton number $Z$ and mass number 
$A$, and $\alpha = e,x$. As shown in Fig.~\ref{fig:brem}, the leading-order 
Feynman diagrams for this process are very similar to those for photo-neutrino 
emission, except that the photon here is linked to the nucleus and thus is off-shell 
(virtual). Both charged-current (CC, $W$-exchange) and neutral-current 
(NC, $Z^0$-exchange) interactions contribute to $\nu_e\bar\nu_e$ pair production, 
while only NC interactions contribute to $\nu_x\bar\nu_x$ pair production. 

In the hot and dense stellar interior, medium effects should be taken into account
for thermal processes. For electron-nucleus bremsstrahlung, an important effect is
electron screening that modifies the Coulomb interaction between the electron and 
the nucleus. The effective screened potential for a single nucleus $(Z,A)$ 
can be written in the momentum space as
\begin{equation}
V_{\rm eff}(|{\bf k}|)=\frac{Zef(|{\bf k}|)}{{\bf k}^2\epsilon(|{\bf k}|)},
\label{eq-vk}
\end{equation}
where ${\bf k}$ is the momentum transfer to the electron,
$e$ is the magnitude of the electron charge, and $\epsilon(|{\bf k}|)$ is the static 
dielectric function that accounts for electron screening. In Eq.~(\ref{eq-vk}),
\begin{align}
f(|{\bf k}|) =\frac{3[\sin(|{\bf k}|r_c)-(|{\bf k}|r_c)\cos(|{\bf k}|r_c)]}{(|{\bf k}|r_c)^3}
\end{align}
is the form factor corresponding to a uniform charge distribution within
the charge radius $r_c$ for the nucleus $(Z,A)$. We consider
$\rho<4.3 \times 10^{11}\ \text{g cm}^{-3}$, for which medium effects on
the charge distribution within a nucleus can be ignored, and take
$r_c =1.15A^{1/3}\ \text{fm}$. In general, the effective potential in
Eq.~(\ref{eq-vk}) cannot be simply applied to all nuclei in the medium. 
This is because ionic correlations can be important and a structure factor 
$S_\Gamma(|{\bf k}|)$ is required to account for these. 
Below we discuss $\epsilon(|{\bf k}|)$ and $S_\Gamma(|{\bf k}|)$ in some detail, 
and then derive the matrix elements and 
the energy-differential rates for neutrino emission from electron-nucleus 
bremsstrahlung. Throughout the paper, we use the natural units where 
the reduced Planck constant $\hbar$ and the speed of light $c$ are set 
to unity. 

\subsection{Static dielectric function $\epsilon(|{\bf k}|)$} 
The hot and dense stellar matter is composed of $e^\pm$ in 
a background of positive ions. For the conditions of interest, 
these ions are simply the bare nuclei. As $e^\pm$ are much lighter and 
thus more mobile than ions, screening of the electron-nucleus Coulomb 
interaction is caused by $e^\pm$. To a good approximation, we can
assume that ions are fixed and discuss how a static electric field is 
screened by $e^\pm$. We first calculate the screening effect by 
generalizing the semiclassical approximation used in Ref.~\cite{Cassola71}
to include both $e^-$ and $e^+$.

In an ideal gas, the equilibrium $e^\pm$ number densities are
\begin{align}
n_{e^\pm}=\frac{2}{(2\pi)^3}\int_0^\infty N_\pm(E)d^3{\bf p}\equiv
\frac{1}{4\pi^3}\int_0^\infty\frac{d^3{\bf p}}{\exp[(E\pm\mu)/(k_BT)]+1},
\end{align}
where $N_\pm(E)$ are the $e^\pm$ occupation numbers at energy $E$,
${\bf p}$ is the corresponding momentum, $\mu$ is the chemical potential, 
$k_B$ is the Boltzmann constant, and $T$ is the temperature. 
For a specific $T$, $\mu$ can be obtained from the net
electron number density $n_e\equiv n_{e^-}-n_{e^+}=\rho/(\mu_em_u)$,
where $\rho$ is the mass density of nuclei associated with the $e^\pm$ gas, 
$m_u$ is the atomic mass unit, and $\mu_e$ is the molecular weight 
per net electron. For the simple case of a neutral uniform 
one-component plasma (OCP), $\mu_e=A/Z$.
When a nucleus $(Z,A)$ is introduced into this OCP, its screened
potential $\phi(r)$ shifts the equilibrium $e^\pm$ number densities at a
distance $r$ to
\begin{align}
n_{e^\pm}'(r)=\frac{1}{4\pi^3}\int_0^\infty\frac{d^3{\bf p}}
{\exp[(E\pm e\phi(r)\pm\mu)/(k_BT)]+1}.
\end{align}
Relative to the initial uniform OCP, the changes in the $e^\pm$ number 
densities to the leading order are
\begin{align}
\delta n_{e^\pm}(r)\equiv n_{e^\pm}'(r)-n_{e^\pm}\approx
\mp\frac{e\phi(r)}{4\pi^3k_BT}\int_0^\infty
\frac{\exp[(E\pm\mu)/(k_BT)]d^3{\bf p}}
{\{\exp[(E\pm\mu)/(k_BT)]+1\}^2}.
\label{eq-dne}
\end{align}

According to Poisson's equation,
\begin{align}
\nabla^2\phi&=e(\delta n_{e^-}-\delta n_{e^+})-Ze\delta({\bf r})\nonumber\\
&\approx\frac{4\alpha\phi(r)}{\pi}\int_0^\infty\left[N_-(E)+N_+(E)\right]
\frac{{\bf p}^2}{E}\left(1+\frac{1}{v^2}\right)d|{\bf p}|-Ze\delta({\bf r}),
\label{eq-po}
\end{align}
where $\alpha\equiv e^2/(4\pi)$ and $v=|{\bf p}|/E$. The approximate result in
Eq.~(\ref{eq-po}) is obtained by using Eq.~(\ref{eq-dne}) and performing 
integration by parts. The solution to Eq.~(\ref{eq-po}) in the momentum 
space is
\begin{align}
V(|{\bf k}|)=\int\phi(r)\exp(-i{\bf k\cdot r})d^3{\bf r}=\frac{Ze}{{\bf k}^2\epsilon(|{\bf k}|)},
\end{align}
where
\begin{align}
\epsilon(|{\bf k}|)\approx 1+\frac{4\alpha}{\pi {\bf k}^2}\int_0^\infty\left[N_-(E)+N_+(E)\right]
\frac{{\bf p}^2}{E}\left(1+\frac{1}{v^2}\right)d|{\bf p}|
\label{eq-eps}
\end{align}
is the static dielectric function.
        
The electron screening effect also affects the propagation of photons in the
$e^\pm$ plasma. For a photon of energy $\omega$ and momentum ${\bf k}$,
the longitudinal component $\Pi_l(\omega,|{\bf k}|)$ of its polarization tensor
to the first order in $\alpha$ \cite{Braaten93} is
\begin{align}
\Pi_l(\omega,|{\bf k}|)=\frac{4\alpha}{\pi}\int_0^\infty\left(\frac{\omega}{v|{\bf k}|}
\ln\frac{\omega+v|{\bf k}|}{\omega-v|{\bf k}|} -1 - 
\frac{\omega^2-{\bf k}^2}{\omega^2-v^2{\bf k}^2}\right)[N_-(E)+N_+(E)]
\frac{{\bf p}^2}{E}d|{\bf p}|.
\label{eq:pi_l1}
\end{align}
The static dielectric function $\epsilon(|{\bf k}|)$ is related to $\Pi_l(\omega,|{\bf k}|)$ 
for $\omega=0$ as $\epsilon(|{\bf k}|)=1-\Pi_l(0,|{\bf k}|)/{\bf k}^2$, which gives the 
same result as Eq.~(\ref{eq-eps}) derived from the semiclassical approximation. 

\begin{figure}[htbp]
\centering
\includegraphics[width=8.cm]{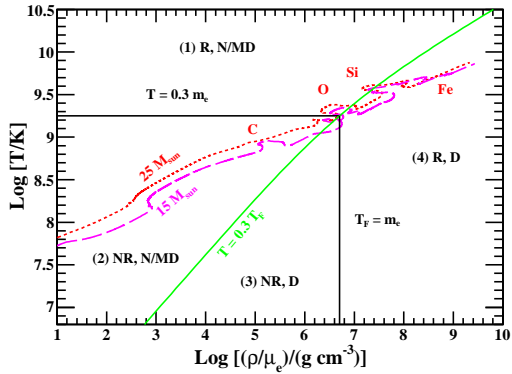}
\caption{Regions of the $(T,\rho/\mu_e)$ space where electrons are 
(1) relativistic and non-degenerate or moderately degenerate (R, N/MD), 
(2) non-relativistic and non-degenerate or moderately degenerate (NR, N/MD), 
(3) non-relativistic and degenerate (NR, D), and (4) relativistic and degenerate 
(R, D), respectively. Also shown are the evolutionary tracks of $T$ and 
$\rho/\mu_e$ at the center for two stars of 15 and $25\,M_\odot$, 
respectively. The labels for the tracks indicate approximately stages of
core C, O, and Si burning and Fe core formation, respectively.}
\label{fig:regions}
\end{figure}

Note that Eq.~(\ref{eq-eps}) applies to all $T$ and $\rho/\mu_e$ and 
is accurate to the first order in $\alpha$ \cite{Braaten93}.
The conditions in a massive star span a wide range of $T$ and 
$\rho/\mu_e$ during its pre-supernova evolution. Figure~\ref{fig:regions} shows
the evolutionary tracks of $T$ and $\rho/\mu_e$ at the center for
two stars of 15 and $25\,M_\odot$, respectively.
The $(T,\rho/\mu_e)$ space can be approximately divided into four
regions: (1) $T>0.3m_e$ and $T>0.3T_F$, where the $e^\pm$ gas
is relativistic and non-degenerate or moderately degenerate (R, N/MD), 
(2) $0.3T_F<T<0.3m_e$, where the gas is non-relativistic and 
non-degenerate or moderately degenerate (NR, N/MD),
(3) $T<0.3T_F$ and $T_F<m_e$, where the gas is non-relativistic and 
degenerate (NR, D), and (4) $T<0.3T_F$ and $T_F>m_e$, where the 
gas is relativistic and degenerate (R, D). Here $m_e$ is the electron 
mass and $T_F$ is the electron Fermi temperature defined as
\begin{align}
T_F\equiv\frac{\sqrt{p_F^2+m_e^2}-m_e}{k_B}= 5.930\times10^9
\left\{\left[1+1.018(\rho_6/\mu_e)^{2/3}\right]^{1/2} -1\right\}\ {\rm K},
\end{align}
where $p_F=(3\pi^2n_e)^{1/3}$ is the electron Fermi momentum,
and $\rho_6$ is $\rho$ in units of $10^6$~g~cm$^{-3}$. 
Figure~\ref{fig:regions} shows that massive stars undergoing core oxygen 
(O) burning encounter conditions at the boundary of the above four regions, 
for which Eq.~(\ref{eq-eps}) should be used to evaluate the static dielectric 
function $\epsilon(|{\bf k}|)$. We have checked that the approximate 
expressions adopted in Refs.~\cite{Itoh83,Itoh87} give the same results as 
Eq.~(\ref{eq-eps}) only when positrons can be ignored [i.e., well
within the (NR, ND), (NR, D), and (R,D) regions in Fig.~\ref{fig:regions}].
We use Eq.~(\ref{eq-eps}) for $\epsilon(|{\bf k}|)$ in our calculations below.

\subsection{Structure factor $S_\Gamma(|{\bf k}|)$}
\label{sec-sf}
For the conditions of interest, $e^\pm$ can always be treated as
in a gas state. They are scattered by the total (screened) Coulomb 
potential generated by all ions. When the temperature is high and/or 
the matter density is low, the ions are also in the gas state and 
each ion can be treated independently. The total rate for electron-nucleus 
scattering is simply the sum over each single ion. However, the dense 
stellar interiors can give rise to condensed states with strong correlations 
among ions. Previous studies \cite{Itoh83,Itoh84a,Itoh84c,Itoh87} 
showed that these ionic correlations
have substantial effects on neutrino emission from electron-nucleus 
bremsstrahlung and, therefore, should be treated properly. 

Correlation effects are described by the structure factor. Consider $N$ 
ions located at ${\bf R}_i$ ($i=1,2,...,N$) in an OCP. The ionic number 
density is $n_I({\bf r}) = \sum_i^N \delta({\bf r} - {\bf R}_i)$. In the static 
case and by the Born approximation, the total electron-nucleus scattering
amplitude is proportional to
\begin{align}
V_{\rm tot}({\bf k}) = V_{\rm eff}(|{\bf k}|)\int n_I({\bf r})\exp(-i{\bf k\cdot r})d^3{\bf r} 
= V_{\rm eff}(|{\bf k}|)\sum_i^N\exp(-i{\bf k\cdot R}_i).
\end{align}
The total scattering rate is proportional to 
$|V_{\rm tot}({\bf k})|^2 = |V_{\rm eff}(|{\bf k}|)|^2 \sum_{ij}
\exp[-i{\bf k\cdot}({\bf R}_i-{\bf R}_j)]$. Taking a time average of the OCP,
we obtain 
\begin{align}
\left\langle|V_{\rm tot}({\bf k})|^2\right\rangle = |V_{\rm eff}(|{\bf k}|)|^2
\left\langle\sum_{ij}\exp[-i{\bf k\cdot}({\bf R}_i-{\bf R}_j)]\right\rangle
\equiv |V_{\rm eff}(|{\bf k}|)|^2NS({\bf k}),
\label{eq-sk}
\end{align}
where $S({\bf k})$ is the static structure factor defined by the second equality. 
For an isotropic system, $S({\bf k})=S(|{\bf k}|)$. As a simple illustration, consider
the gas state in which the correlations among ions are weak due to random 
thermal motion at high temperature and/or the feeble interaction between 
two distant ions at low density. For this case, $S(|{\bf k}|)\approx 1$ as only 
those terms with $i=j$ in the sum in Eq.~(\ref{eq-sk}) are not averaged out.
Consequently, the total rate for electron-nucleus scattering in this case is
just $N$ times the rate for a single nucleus. 
 
In general, the ionic state of an OCP can be characterized by the parameter
\begin{align}
\Gamma\equiv\frac{Z^2e^2}{a_I k_B T} = 0.2275 \frac{Z^2}{T_8} 
\left(\frac{\rho_6}{A}\right)^{1/3}, 
\label{eq:TGamma} 
\end{align}
where $T_8$ is $T$ in units of $10^8$~K, $a_I=[3/(4\pi\bar n_I)]^{1/3}$ 
is the ion-sphere radius, and $\bar n_I=\rho/(Am_u)$ is the mean ion
number density. As can be seen from its definition, $\Gamma$ measures 
the Coulomb interaction energy between two nearby ions relative to their
thermal energy. The gas, liquid, and crystal lattice states correspond to
$\Gamma\ll 1$, $1\lesssim\Gamma\lesssim 180$, and $\Gamma>180$, 
respectively. The structure factor is rather complex for the liquid and 
crystal lattice states. To indicate its dependence on $\Gamma$, we 
denote it as $S_\Gamma(|{\bf k}|)$.

\begin{figure}[htbp]
\centering
\includegraphics[width=8.cm]{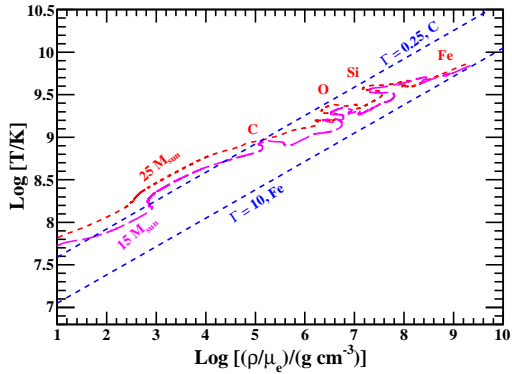}
\caption{Contours in the $(T,\rho/\mu_e)$ space corresponding to 
$\Gamma=0.25$ for an OCP composed of $^{12}$C and $\Gamma=10$ for 
an OCP composed of $^{56}$Fe. Also shown are the same two stellar
evolutionary tracks as in Fig.~\ref{fig:regions}.}
\label{fig:gamma}
\end{figure}

Figure~\ref{fig:gamma} shows contours of $\Gamma$ for an OCP 
composed of $^{12}$C or $^{56}$Fe along with the evolutionary tracks 
of $T$ and $\rho/\mu_e$ at the center for two stars of 15 and $25\,M_\odot$,
respectively. It can be seen that ions are in the gas or liquid state 
($\Gamma\lesssim 10$) during 
the pre-supernova evolution of massive stars. An analytic fit to the structure 
factor $S_\Gamma(|{\bf k}|)$ for an OCP was provided by Ref.~\cite{Young91} 
based on the results calculated from the modified hypernetted-chain 
equation for $0.1\leq\Gamma\leq 225$ \cite{Rogers83}. Although the fit was obtained 
for $1\leq\Gamma\leq 225$, we find that its extension to $\Gamma<1$ remains
a good approximation to the results calculated in Ref.~\cite{Rogers83} even for 
$\Gamma=0.1$. It also has the correct asymptotic behavior 
$S_\Gamma(|{\bf k}|)\to 1$ for $\Gamma\to 0$. Therefore, this fit is sufficient for
our discussion on the spectra and rates for neutrino emission from
electron-nucleus bremsstrahlung during the pre-supernova evolution of 
massive stars.

The crystal lattice state may be reached for $\Gamma\gtrsim 210$ during 
the cooling of dense stars \cite{Ichimaru83}. The structure factor in this regime is
needed for a general discussion of the conditions under which neutrino 
energy loss is dominated by electron-nucleus bremsstrahlung. Here new
effects associated with the thermal motion of ions, the band structure of 
electrons \cite{Pethick97}, and multi-phonon processes \cite{Kaminker98}
must be taken into account. We follow the discussion in Ref.~\cite{Kaminker98} 
to calculate $S_\Gamma(|{\bf k}|)$ for the crystal lattice state and refer
readers to that work for details. The resulting prescription gives 
similar values for $S_\Gamma(|{\bf k}|)$ to those from the fit in Ref.~\cite{Young91} 
for $100\lesssim\Gamma\lesssim 225$ \cite{Kaminker98}. In our calculations,
we adopt the fit in Ref.~\cite{Young91} for $\Gamma\leq 180$ and the
prescription in Ref.~\cite{Kaminker98} for $\Gamma>180$. 
                                             
\subsection{Matrix elements and energy-differential rates}
We now calculate the matrix elements for $\nu_e$ and $\bar\nu_e$
emission from electron-nucleus bremsstrahlung. We first ignore ionic 
correlations. The amplitude for the $Z^0$-exchange diagrams can be 
written in standard notation of the electroweak theory as
\begin{align}
i{\cal M}_{Z^0} = &-\frac{iZe^2G_F^2}{2\sqrt{2}} 
\frac{f(|{\bf k}|)}{{\bf k}^2\epsilon(|{\bf k}|)}
\left[\bar u_e(p^\prime)\gamma^\alpha(a+b\gamma^5)(\slashed p + \slashed k - m_e)^{-1} 
\gamma^0 u_e(p)\bar{u}_\nu(q) \gamma_\alpha(1-\gamma^5)v_{\bar \nu}(q^\prime)\right.\nonumber \\
&\left.+ \bar u_e(p^\prime) \gamma^0 (\slashed p^\prime - \slashed k - m_e)^{-1} \gamma^\alpha
(a+b\gamma^5) u_e(p)\bar{u}_\nu(q) \gamma_\alpha(1-\gamma^5)v_{\bar \nu}(q^\prime)\right],
\label{eq:amp_Z}
\end{align}
where $G_F$ is the Fermi coupling constant, $a=-1+4s_W^2$, 
$s_W \equiv \sin\theta_W$ with $\theta_W$ being the Weinberg angle, 
$b=1$, $\gamma^\alpha\ (\alpha=0,1,2,3)$ and $\gamma^5$
refer to the Dirac gamma matrices, $u$ and $v$ are spinors, and the 
four-momenta $k$, $p$, $p'$, $q$, and $q'$ are as labeled in 
Fig.~\ref{fig:brem}. Note that the four-momentum for the virtual photon 
is $k=(0,{\bf k})$.

The amplitude for the $W$-exchange diagrams can be arranged to
have a similar structure to that for the $Z^0$-exchange diagrams 
via Fierz transformations. The total amplitude for both types of 
diagrams is
\begin{align}
i{\cal M} = & i({\cal M}_{Z^0} + {\cal M}_W)\nonumber \\
=& -\frac{iZe^2 G_F}{\sqrt{2}}\frac{f(|{\bf k}|)}{{\bf k}^2\epsilon(|{\bf k}|)}
\left[\bar u_e(p^\prime)\gamma^\alpha(C_V-C_A\gamma^5)
(\slashed p + \slashed k - m_e)^{-1} \gamma_0 u_e(p)
\bar{u}_\nu(q) \gamma_\alpha(1-\gamma^5)v_{\bar \nu}(q^\prime)\right.\nonumber \\
&+\left.\bar u_e(p^\prime) \gamma_0 (\slashed p^\prime - \slashed k - m_e)^{-1} 
\gamma^\alpha(C_V-C_A\gamma^5) u_e(p)\bar{u}_\nu(q) \gamma_\alpha
(1-\gamma^5)v_{\bar \nu}(q^\prime)\right],
\label{amp_brem}
\end{align}
where $C_V = (1+4s_W^2)/2$ and $C_A = 1/2$. After averaging over 
the fermion spins in the initial state and summing over those in the final 
state, we obtain the effective squared matrix element
\begin{align}
\vert{\cal M}\vert^2_{\rm eff} =&\frac{Z^2e^4G_F^2}{4}
\frac{[f(|{\bf k}|)]^2}{[{\bf k}^2\epsilon(|{\bf k}|)]^2}\nonumber\\
&\times\mbox{Tr}\left\{(\slashed p^\prime+m_e)
\left[\gamma^\alpha(C_V-C_A\gamma^5) \frac{{\slashed Q}_1+m_e}{\beta_1} 
\slashed \epsilon_B+\slashed \epsilon_B \frac{{\slashed Q}_2+m_e}{\beta_2} 
\gamma^\alpha(C_V-C_A\gamma^5)\right](\slashed p+m_e)\right.\nonumber \\
&\times\left.\left[(C_V + C_A \gamma^5) \gamma^\beta 
\frac{{\slashed Q}_2+m_e}{\beta_2} \slashed\epsilon_B
+\slashed \epsilon_B\frac{{\slashed Q}_1+m_e}{\beta_1}(C_V+C_A\gamma^5) 
\gamma^\beta\right]\right\} \nonumber \\
& \times \mbox{Tr}\left[ \slashed q \gamma_\alpha(1-\gamma^5)\slashed q^\prime 
(1+\gamma^5)\gamma_\beta\right], 
\label{eq-amp2}
\end{align}
where $Q_1 \equiv p+k$, $Q_2 \equiv p^\prime - k$, 
$\beta_1 \equiv 2k\cdot p - {\bf k}^2$, and
$\beta_2 \equiv -2k\cdot p^\prime - {\bf k}^2$. 
Note that we have defined an artificial polarization four-vector 
$\epsilon_B = (1,0,0,0)$ to make the result similar to that for 
photo-neutrino emission \cite{Dicus72,Dutta03}. With the definition of $I_i^B\ (i=1,2,3)$ in 
Appendix~\ref{sec:app_sqamp}, Eq.~(\ref{eq-amp2}) can be rewritten as
\begin{align}
\vert {\cal M} \vert^2_{\rm eff} = 4 Z^2e^4G_F^2 
\frac{[f(|{\bf k}|)]^2}{[{\bf k}^2\epsilon(|{\bf k}|)]^2}\left[(C_V^2+C_A^2)I_1^B
+ (C_V^2-C_A^2)I_2^B + C_V C_A I_3^B\right].
\label{eq-m2eff}
\end{align}

\subsubsection{OCP}  
Including the structure factor $S_\Gamma(|{\bf k}|)$ to account for ionic 
correlations and integrating over the phase space of the initial and 
final states, we obtain the energy-differential rate per unit volume 
for $\nu_e$ emission from electron-nucleus bremsstrahlung in an OCP as   
\begin{align}
F_{\nu_e}(E_\nu) =&\frac{\rho}{Am_u}\int \frac{2d^3{\bf p}}{2E(2\pi)^3} N_-(E) 
\int\frac{d^3{\bf k}}{(2\pi)^3} S_\Gamma(|{\bf k}|)
\int\frac{d^3{\bf p}^\prime}{2E^\prime(2\pi)^3}\left[1-N_-(E^\prime)\right]
\nonumber\\ 
&\times\int\frac{d^3{\bf q}'}{2E_\nu'(2\pi)^3}
\int\frac{E_\nu^2d\Omega_{\bf q}}{2E_\nu(2\pi)^3}
(2\pi)^4\delta(p+k-p'-q-q')\vert {\cal M} \vert^2_{\rm eff}\nonumber\\
=&\frac{\rho}{Am_u}\int\frac{2d^3{\bf p}}{2E(2\pi)^3} N_-(E) 
\int\frac{d^3{\bf k}}{(2\pi)^3} S_\Gamma(|{\bf k}|)
\int\frac{d^3{\bf p}^\prime}{2E^\prime(2\pi)^3}\left[1-N_-(E^\prime)\right] 
\frac{\int d\varphi \vert {\cal M} \vert^2_{\rm eff}}{16\pi^2 \vert {\bf P}\vert},
\label{eq:brem_spec}     
\end{align} 
where ${\bf P} \equiv {\bf p} + {\bf k} -{\bf p^\prime}$, and $\Omega_{\bf q}$ 
is the solid angle for the $\nu_e$ momentum ${\bf q}$ with $\varphi$ being
the azimuth angle around ${\bf P}$. Note that the $\delta$ function in the 
above equation is disposed of by integration over the $\bar\nu_e$ momentum 
${\bf q}'$ and the polar angle of ${\bf q}$ with respect to ${\bf P}$.
For $\nu_x$, only $Z^0$-exchange diagrams contribute, and the 
corresponding differential rate $F_{\nu_x}(E_\nu)$ is obtained by replacing 
$C_V$ and $C_A$ in $|{\cal M}|_{\rm eff}^2$ [see Eq.~(\ref{eq-m2eff})] with 
$a$ and $-b$ as defined for Eq.~(\ref{eq:amp_Z}), respectively.

Note that only the neutrino four-momentum $q$ shows up explicitly in the 
expressions for $I_i^B\ (i=1,2,3)$ [see Eqs.~(\ref{eq:IB_1})--(\ref{eq:brem_amp5})] 
as the antineutrino four-momentum $q^\prime$ has been evaluated by applying 
conservation of energy and momentum. If $q^\prime$ is kept instead of $q$,
the new expressions have the same form but with $q^\prime$ replacing
$q$ and an opposite sign for $I_3^B$. In other words, $I_1^B$ and $I_2^B$
are symmetric while $I_3^B$ is antisymmetric under the exchange of $q$ and $q'$.
Therefore, a simple way to obtain $F_{\bar\nu_\alpha}(E_\nu)$ is to change the 
sign of the contribution from the $I_3^B$ term in $F_{\nu_\alpha}(E_\nu)$. This 
sign change makes the neutrino and antineutrino spectra somewhat different. 
However, when integrated over the $\nu_\alpha$ and $\bar\nu_\alpha$ phase 
space to obtain the total rates of emission $R_{\nu_\alpha}$ 
($R_{\bar\nu_\alpha}$) and energy loss $Q_{\nu_\alpha}$ 
($Q_{\bar\nu_\alpha}$) in $\nu_\alpha$ ($\bar\nu_\alpha$), 
the $I_3^B$ term does not contribute. The contributions from the $I_1^B$ 
and $I_2^B$ terms always ensure that $R_{\nu_\alpha} = R_{\bar\nu_\alpha}$ 
and $Q_{\nu_\alpha} = Q_{\bar\nu_\alpha}$. 

Finally, positrons can also scatter on nuclei to produce 
$\nu_\alpha\bar\nu_\alpha$ pairs. For the same incoming and 
outgoing four-momenta $p$ and $p'$, respectively, the amplitudes of
positron-nucleus bremsstrahlung can be obtained by interchanging 
$p$ and $-p'$ in the results for electron-nucleus bremsstrahlung. 
The terms $I_1^B$ and $I_2^B$ are symmetric while $I_3^B$ is 
antisymmetric under this interchange. Therefore, the differential 
rates for positron-nucleus bremsstrahlung can be obtained by 
replacing $N_-(E)$ [$N_-(E')$] with $N_+(E)$ [$N_+(E')$] and changing
the sign of $I_3^B$ in Eq.~(\ref{eq:brem_spec}). We include
the contributions from both electrons and positrons to bremsstrahlung
neutrino emission in our numerical results.

\subsubsection{Multi-Component Plasma} 
Stellar matter typically consists of more than one nuclear species, and
thus corresponds to a multi-component plasma (MCP). To extend our results to this case, we
follow Ref.~\cite{Itoh04} and treat the different nuclear components 
independently. The state of nuclei $(Z_j, A_j)$ is determined by 
\begin{align}
\Gamma_j = \frac{Z_j^2e^2}{a_j k_B T} = 0.2275 \frac{Z_j^{5/3}}{T_8} 
\left(\rho_6\sum_i \frac{x_i Z_i}{A_i}\right)^{1/3}, 
\end{align}
where $a_j$ is defined by
\begin{align}
\frac{4\pi}{3}a_j^3\sum_i \frac{x_i\rho}{A_im_u} Z_i = Z_j, 
\end{align}
and $x_i$ is the mass fraction of nuclei $(Z_i, A_i)$.  For an OCP,
$\Gamma_j$ and $a_j$ reduce to $\Gamma$ [see Eq.~(\ref{eq:TGamma})] and $a_I$, respectively.
For the $j$th component, the same structure factor 
$S_{\Gamma_j}(|\bf k|)$ as for an OCP is used to account for 
ionic correlations. Summing the contributions from each component
incoherently, we can generalize the energy-differential rate in
Eq.~(\ref{eq:brem_spec}) as
\begin{align}
F_{\nu_e}(E_\nu) 
=\sum_j \frac{x_j\rho}{A_jm_u}\int\frac{2d^3{\bf p}}{2E(2\pi)^3} N_-(E) 
\int\frac{d^3{\bf k}}{(2\pi)^3} S_{\Gamma_j}(|{\bf k}|)
\int\frac{d^3{\bf p}^\prime}{2E^\prime(2\pi)^3}\left[1-N_-(E^\prime)\right] 
\frac{\int d\varphi \vert {\cal M}_j\vert^2_{\rm eff}}{16\pi^2 \vert {\bf P}\vert},
\label{eq:brem_spec_mcp}     
\end{align}
where $\vert {\cal M}_j \vert^2_{\rm eff}$ is given by Eq.~(\ref{eq-m2eff})
with $Z$ replaced by $Z_j$.

The above approximate method of treating an MCP has some limitation
\cite{Sawyer05}. However, for the neutrino energy range of interest,
$E_\nu \gtrsim 0.1$~MeV, the results based on this method are consistent 
with those from simulations based on molecular dynamics \cite{Caballero06}. 
The same method has also been adopted to treat neutrino-nucleus 
scattering during stellar core collapse \cite{Marek05}.

\section{Comparison of spectra and rates for thermal neutrino emission processes}   
\label{sec:spec}
The energy-differential rate in Eq.~(\ref{eq:brem_spec}) is highly non-trivial
to calculate. Our numerical computation proceeds as follows. We pick ${\bf p}$ 
as the $z$-direction and define a coordinate system. By specifying 
$|{\bf p}|$, ${\bf k}$, and ${\bf p}'$, we fix ${\bf P}={\bf p}+{\bf k}-{\bf p}'$ and 
the polar angle of ${\bf q}$ with respect to ${\bf P}$ (through energy and 
momentum conservation). By further specifying the azimuthal angle $\varphi$
of ${\bf q}$ around ${\bf P}$, all the vectors involved in the effective squared 
matrix element $|{\cal M}|_{\rm eff}^2$ are fixed. Therefore, the 
energy-differential rate is an eight-dimensional integral over $\varphi$, 
${\bf p}'$, ${\bf k}$, and $|{\bf p}|$. We use the Vegas Monte Carlo algorithm 
encoded in the {\footnotesize CUBA} library \cite{Cuba2005} to evaluate all the 
multidimensional integrals in this work.

\subsection{Effects of ionic correlations on bremsstrahlung neutrino emission}
\begin{figure}[htp]
\centering
\includegraphics[width=0.5\textwidth]{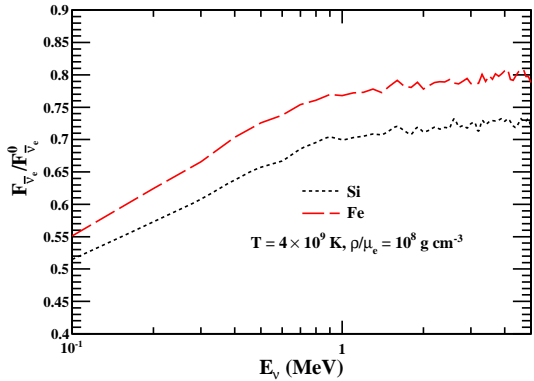}   
\caption{Effects of ionic correlations on bremsstrahlung neutrino emission.
The ratio of the rate $F_{\bar\nu_e}$ with ionic correlations to the rate 
$F_{\bar\nu_e}^0$ without such correlations is shown as a function of 
$\bar\nu_e$ energy $E_\nu$ for an OCP composed of $^{28}$Si or
$^{56}$Fe, respectively.}  
\label{fig:spec_ionic} 
\end{figure}

As discussed in Sec.~\ref{sec-sf}, ionic correlations complicate the calculation
of the energy-differential rate for bremsstrahlung neutrino emission.
Taking $T = 4\times 10^9$~K and $\rho/\mu_e = 10^8$~g~cm$^{-3}$,
we show in Fig. \ref{fig:spec_ionic} the ratio of the rate $F_{\bar\nu_e}$
with ionic correlations to the rate $F_{\bar\nu_e}^0$ without such correlations 
as a function of $\bar\nu_e$ energy $E_\nu$ for an OCP composed of 
$^{28}$Si or $^{56}$Fe with $\Gamma\approx 2.2$ or 6, respectively. 
It can be seen that ionic correlations reduce the energy-differential rate
by a factor of $\sim 2$ at low energies and by a factor of $\sim 1.3$ at high
energies. This result is not sensitive to the composition of the OCP and
we have checked that it holds true generally for $\Gamma \lesssim 10$, which
is relevant for massive stars during their pre-supernova evolution
(see Fig. \ref{fig:gamma}). At $E_\nu\geq 1.8$~MeV, for which $\bar\nu_e$ 
can be detected through capture on protons, $F_{\bar\nu_e}/F_{\bar\nu_e}^0$
varies very slowly and is close to 0.7.

\subsection{Energy-differential rates}
As mentioned in the Introduction, energy-differential rates for $e^\pm$ pair 
annihilation, plasmon decay, and photo-neutrino emission have been studied 
in detail by previous works 
\cite{Odrzywolek04,Dutta03,Ratkovi03,Misiaszek05,Odrzywolek07,Kato15,Patton15}.
We are not aware of a detailed discussion of the energy-differential rate
for bremsstrahlung neutrino emission in the literature. We now discuss this
in comparison with the other thermal emission processes listed above. 
We follow the standard procedures to calculate the energy-differential rates for 
$e^\pm$ annihilation \cite{Odrzywolek04}, plasmon decay 
\cite{Ratkovi03,Odrzywolek07}, and photo-neutrino emission \cite{Dutta03}.
Although details are not presented here, we have used different expressions 
for the squared amplitudes from those in the literature by enforcing energy and
momentum conservation in different ways and we have adopted different 
integration procedures. Therefore, our results for these three processes 
provide an independent check on the previous results.

\begin{figure*}
\centering
\includegraphics[width=0.45\textwidth]{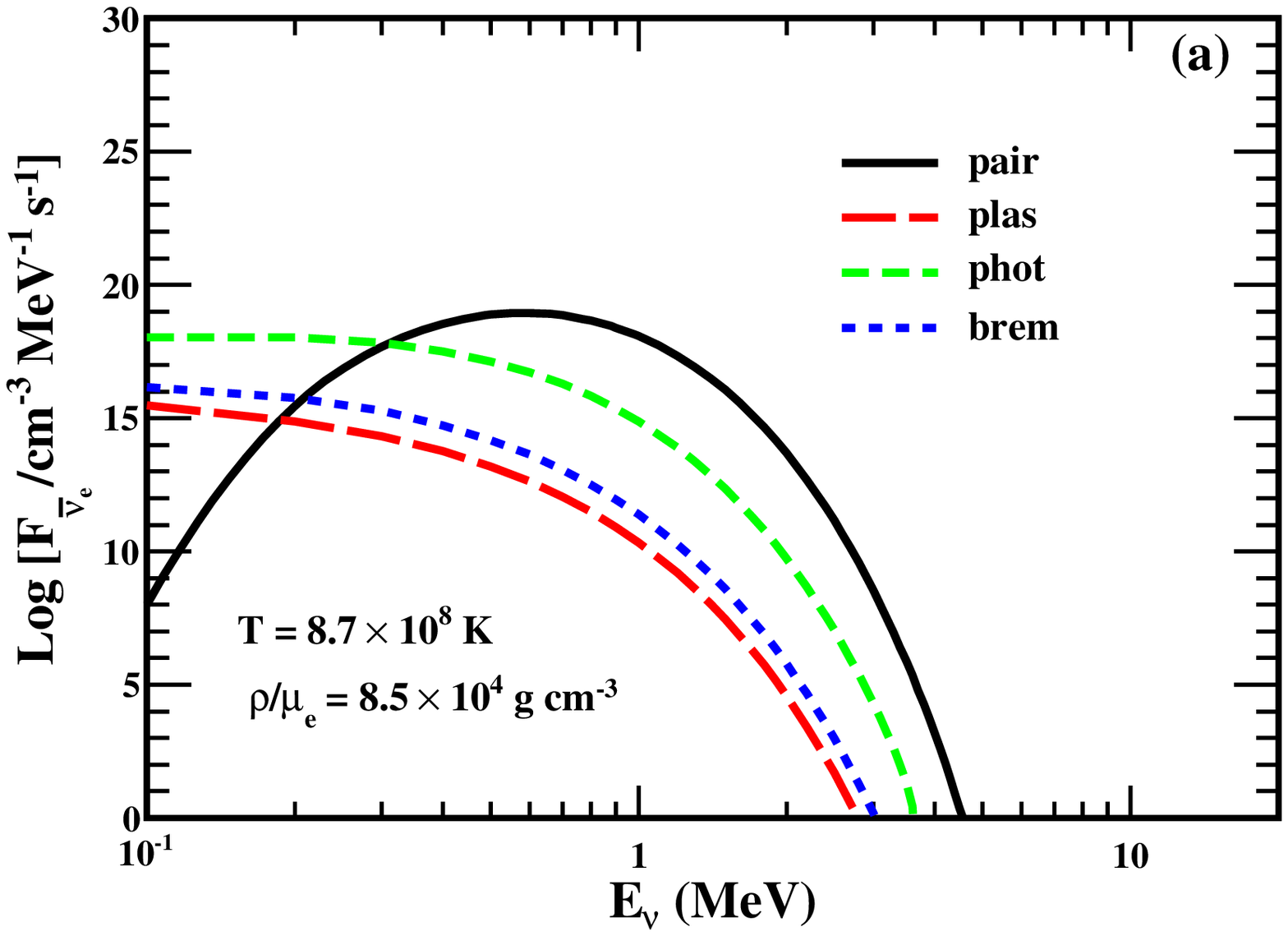} 
\includegraphics[width=0.45\textwidth]{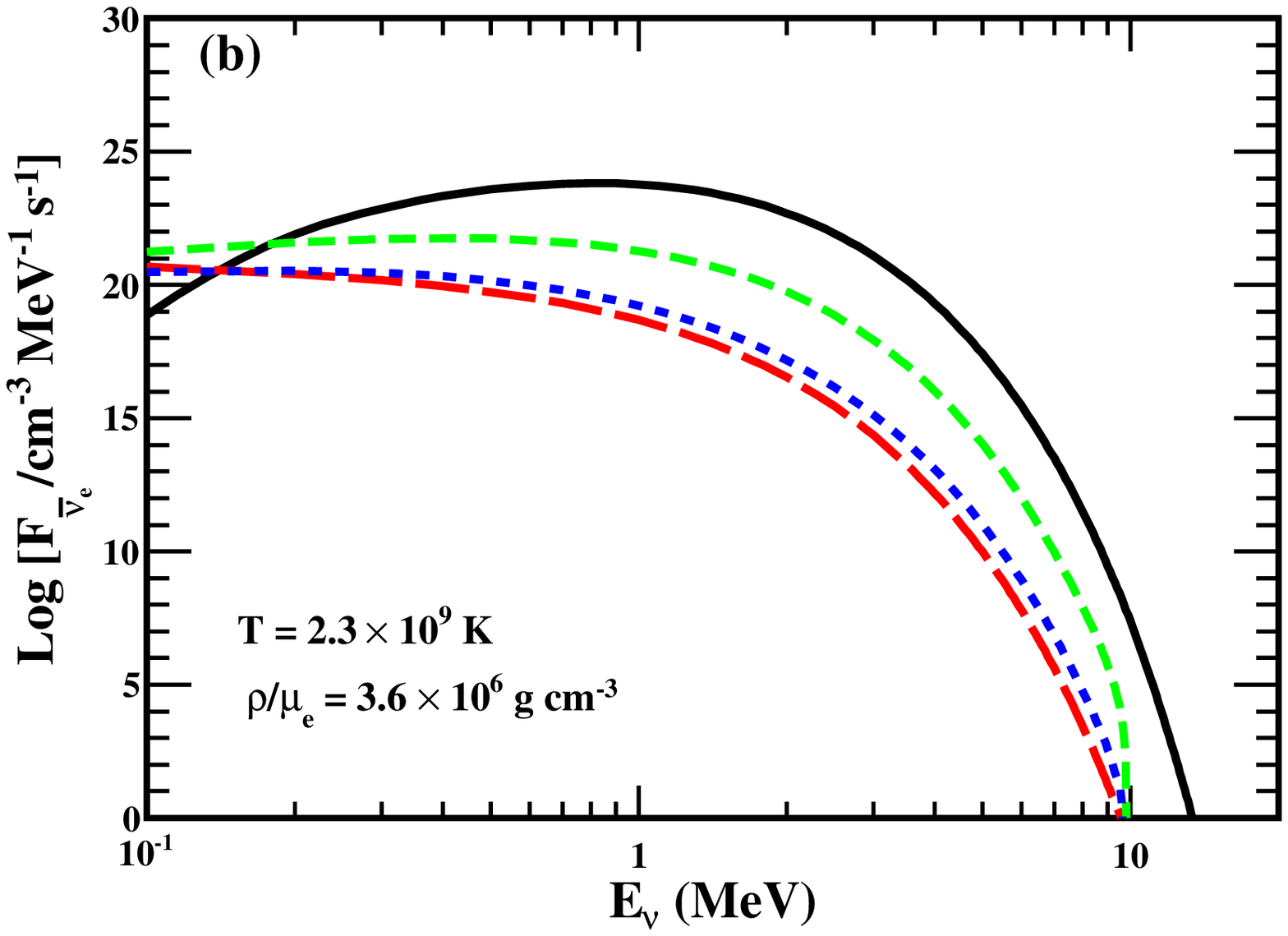}  \\
\includegraphics[width=0.45\textwidth]{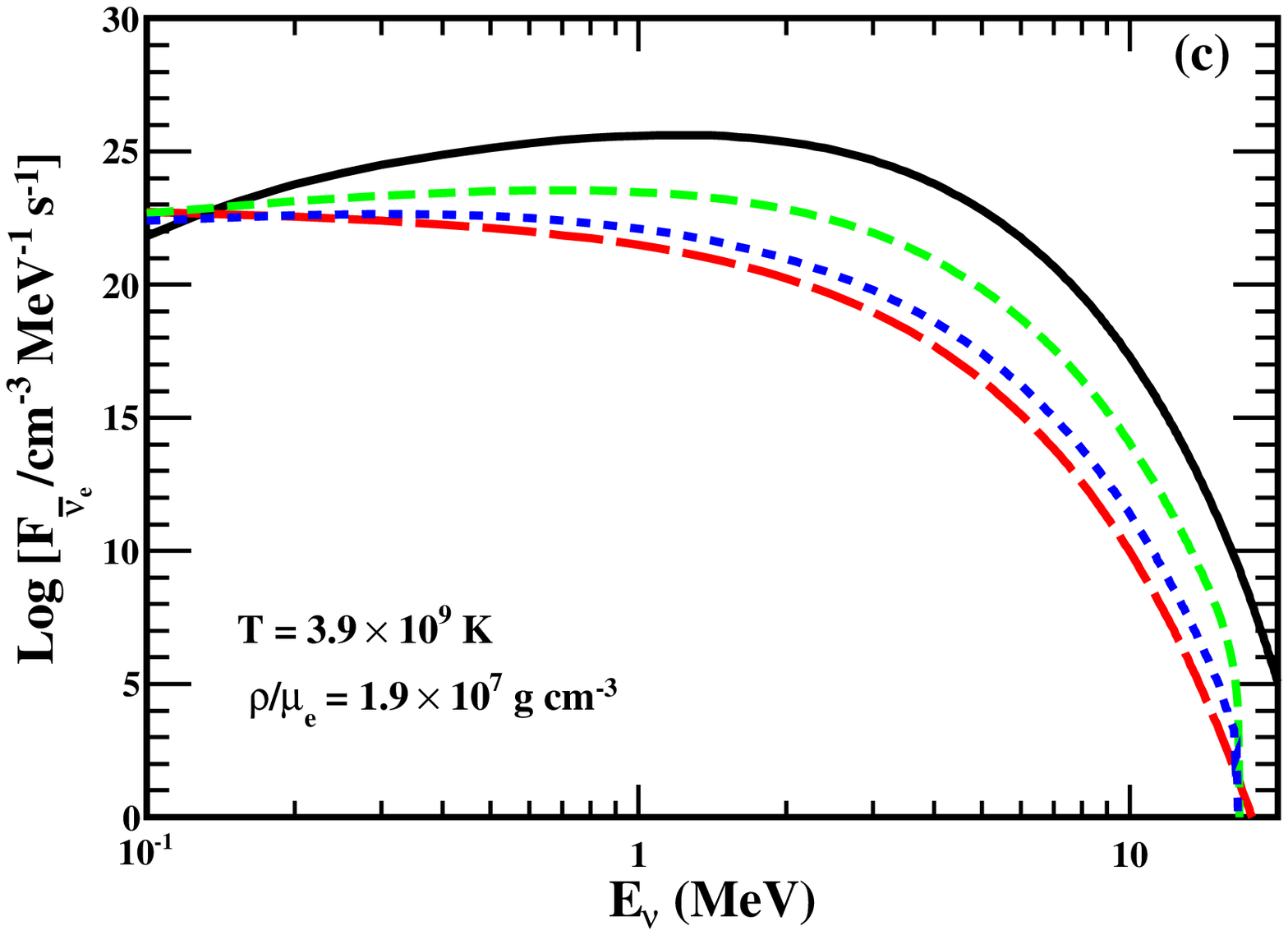} 
\includegraphics[width=0.45\textwidth]{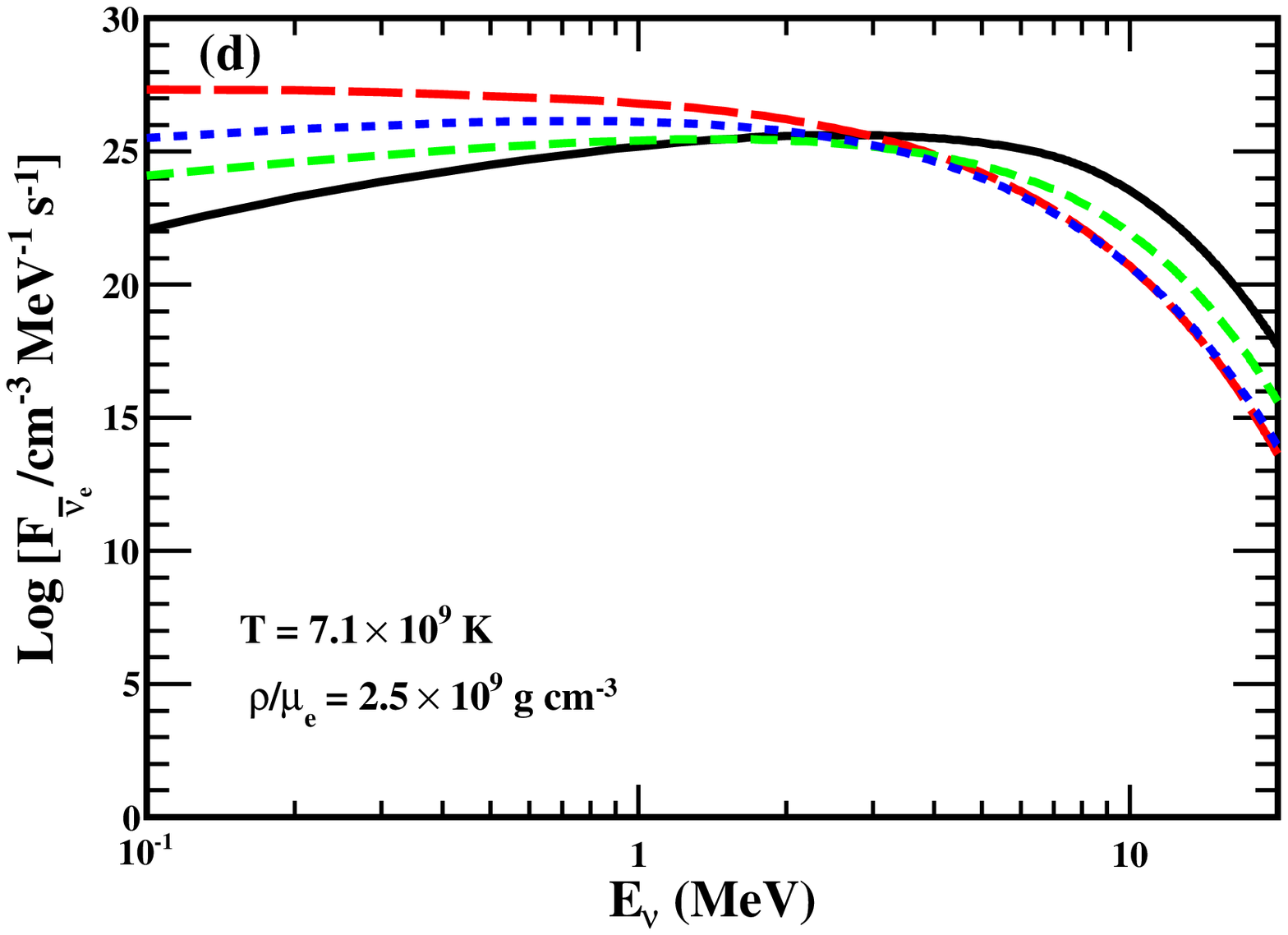}   
\caption{Comparison of energy-differential rates $F_{\bar\nu_e}$ 
as functions of $\bar\nu_e$ energy $E_\nu$ for 
$e^\pm$ pair annihilation (``pair''),
plasmon decay (``plas''), photo-neutrino emission (``phot''), and
bremsstrahlung neutrino emission (``brem''). The conditions indicated
are representative of massive stellar cores (a) during C burning,
(b) at O depletion, (c) at Si depletion, and (d) immediately prior
to collapse.}  
\label{fig:spec}
\end{figure*}

For specific numerical examples, we consider four sets of temperature and 
density $(T_9,\rho_7/\mu_e)=(0.87,8.5\times 10^{-3})$, $(2.3,0.36)$, $(3.9,1.9)$, 
$(7.1,2.5\times10^2)$, which are representative of massive stellar cores 
during C burning, at O depletion, at silicon (Si) depletion, and 
immediately prior to collapse, respectively \cite{massivestar}. 
Here $T_9$ is $T$ in units of $10^9$~K and $\rho_7$ is $\rho$
in units of $10^7$~g~cm$^{-3}$. For calculating the rates for bremsstrahlung
neutrino emission, we simply assume an OCP composed of $^{16}$O, $^{28}$Si,  
$^{56}$Fe, and $^{56}$Fe, respectively, which approximately corresponds 
to the composition for the selected stages of stellar evolution. Contributions 
from other coexisting nuclei can be included in a straightforward manner as shown 
in Eq.~(\ref{eq:brem_spec_mcp}). Because $\Gamma_j$ does not vary much
over the typical composition, $S_{\Gamma_j}$ has similar effects on neutrino 
emission for different components, and the contribution from each component 
is approximately proportional to $x_jZ_j^2/A_j$. 
As the total mass fraction of the subdominant nuclei 
is typically $\lesssim 20\%$, we find that a simple OCP treatment based on
the dominant species introduces errors only at the level of $\sim 10\%$.
The energy-differential rates $F_{\bar\nu_e}$ in units of 
cm$^{-3}$~MeV$^{-1}$~s$^{-1}$ for $\bar\nu_e$ emission from the above 
four processes are shown as functions of 
$\bar\nu_e$ energy $E_\nu$ in Fig.~\ref{fig:spec}.
It can be seen that the differential rate for
$e^\pm$ pair annihilation always dominates at high energies, 
while at low energies, the rates for the other processes become 
comparable or take over. Similar to plasmon decay and 
photo-neutrino emission, bremsstrahlung mostly produces
sub-MeV neutrinos during the pre-supernova evolution of
massive stars. It is interesting to note that bremsstrahlung and plasmon decay have similar spectral shapes, most likely due to similar phase spaces for the outgoing particles. The comparison of $F_{\nu_e}$ for the thermal emission
processes is very similar to that of $F_{\bar\nu_e}$ and, therefore,
is not shown here. 
 
Partly because of the conversion of $e^\pm$ rest
mass into neutrino energy, pair annihilation always produces
the highest average neutrino energy. For example, the average 
$\bar\nu_e$ energy from pair annihilation in a non-degenerate 
and non-relativistic gas of $e^\pm$ can be estimated as 
$\langle E_{\bar\nu_e}\rangle\sim m_e + 3k_BT$. 
In general, the average neutrino energies for different thermal 
emission processes are nontrivial functions of both temperature 
and density. Table~\ref{tab:nu} gives the average 
$\bar\nu_\alpha$ energies $\langle E_{\bar\nu_\alpha}\rangle$
for these processes along with the corresponding net emission 
rates $R_{\bar\nu_\alpha}$ in units of cm$^{-3}$~s$^{-1}$ at the 
four selected stages of stellar evolution. It can be seen 
that both $R_{\bar\nu_\alpha}$ and 
$\langle E_{\bar\nu_\alpha}\rangle$ generally increase for all 
the processes as the star evolves. 
However, the net emission rates for pair annihilation are 
severely suppressed immediately prior to core 
collapse because of strong electron degeneracy. This allows
other processes to compete for energy loss.
Table~\ref{tab:nu} shows that 
$\langle E_{\bar\nu_e}\rangle=\langle E_{\bar\nu_x}\rangle$ 
for plasmon decay, but in general $\langle E_{\bar\nu_e}\rangle$ 
is slightly lower than $\langle E_{\bar\nu_x}\rangle$ for the other 
processes. We note that
$\langle E_{\nu_\alpha}\rangle=\langle E_{\bar\nu_\alpha}\rangle$
for all thermal emission processes.

\begin{table*}[htbp]
\renewcommand\arraystretch{1.2}
\caption{Comparison of rates and characteristic energies for thermal
neutrino emission processes. Four sets of conditions representative
of massive stellar cores during C burning, at O depletion, at Si depletion,
and immediately prior to collapse are chosen. For each case,
results for $e^\pm$ pair annihilation, plasmon decay, photo-neutrino 
emission, and bremsstrahlung neutrino emission are given in four
consecutive rows. $R_{\bar\nu_{e,x}}$ and $\langle E_{\bar\nu_{e,x}}\rangle$ 
are the net emission rates and average energies of $\bar\nu_{e,x}$, respectively. 
$R^>_{\bar\nu_{e,x}}$ and $\bar E^>_{\bar\nu_{e,x}}$ are the net rates and 
average energies for those $\bar\nu_{e,x}$ with energy above 
$E_{\rm th}\approx 1.8$ MeV. $\bar E_{\bar\nu_{e,x}}^{\rm det}$ is an 
effective energy for $\bar\nu_e$ detection, as defined in Eq.~(\ref{eq:Edet}).
$R_{\bar\nu_{e,x}}^>$ are in units of cm$^{-3}$~s$^{-1}$, 
and the characteristic energies $\langle E_{\bar\nu_{e,x}}\rangle$,
$\bar E^>_{\bar\nu_{e,x}}$, and $\bar E_{\bar\nu_{e,x}}^{\rm det}$
are in units of MeV.}
\label{tab:nu} 
\begin{center}
\begin{tabular}{p{3.cm}<{\centering}p{2.5cm}<{\centering}p{2.5cm}
<{\centering}p{2.5cm}<{\centering}p{2.5cm}<{\centering}p{2.5cm}<{\centering}}
\hline
\hline
$(T_9,\rho_7/\mu_e)$& $\log(R_{\bar\nu_{e,x}})$ 
& $\langle E_{\bar\nu_{e,x}}\rangle$ & $\log(R^>_{\bar\nu_{e,x}})$ 
& $\bar E^>_{\bar\nu_{e,x}}$ & $\bar E_{\bar\nu_{e,x}}^{\rm det}$ \\
 $(0.87,8.5\times 10^{-3})$ & 18.57, 17.22 & 0.648, 0.684 & 13.60, 12.58 & 1.892, 1.893 & 1.886, 1.888 \\   
    & 15.05, 12.24 & 0.061, 0.061 & 4.59, 1.79     & 1.877, 1.877 & 1.872, 1.872  \\  
     & 17.54, 17.15 & 0.227, 0.226 & 9.69, 9.29     & 1.885, 1.895 & 1.879, 1.879   \\ 
     & 15.44, 14.96 & 0.131, 0.136 & 5.78, 5.58     & 1.886, 1.898 & 1.876, 1.877  \\       
   $(2.3,0.36)$ & 23.79, 22.86 & 1.006, 1.089 & 22.45, 21.66 & 2.088, 2.092 & 2.118, 2.127  \\   
    & 20.21, 17.40 & 0.176, 0.176 & 16.27, 13.46 & 1.993, 1.993 & 2.019, 2.019  \\  
     & 21.64, 21.13 & 0.607, 0.620 & 19.49, 19.01 & 2.036, 2.032 & 2.072, 2.073  \\   
     & 20.20, 19.69 & 0.348, 0.376 & 16.92, 16.67 & 2.025, 2.017 & 2.036, 2.046 \\
   $(3.9,1.9)$ & 25.82, 25.04 & 1.524, 1.630 & 25.30, 24.58 & 2.372, 2.399 & 2.488, 2.510  \\ 
     & 22.35, 19.54 & 0.314, 0.314 & 20.03, 17.22 & 2.139, 2.139 & 2.199, 2.199  \\  
      & 23.66, 23.08 & 1.040, 1.089 & 22.70, 22.18 & 2.281, 2.295 & 2.363, 2.373 \\   
      & 22.54, 21.97 & 0.592, 0.661 & 20.77, 20.43 & 2.187, 2.201 & 2.234, 2.279 \\ 
 $(7.1,2.5\times10^2)$ & 26.24, 25.55 & 3.558, 4.150 & 26.17, 25.50 & 3.953, 4.472 & 4.387, 4.922  \\   
   & 27.23, 24.42 & 0.765, 0.765 & 26.16, 23.35 & 2.457, 2.457 & 2.620, 2.620 \\    
      & 25.91, 25.25 & 2.177, 2.404 & 25.65, 25.03 & 3.064, 3.231 & 3.379, 3.603  \\  
      & 26.39, 25.73 & 1.322, 1.413 & 25.77, 25.17 & 2.610, 2.632 & 2.758, 2.848   \\
\hline 
\hline
\end{tabular}
\end{center}
\label{default}
\end{table*}%
 
We now consider the potential detection of neutrinos from massive
stars during their pre-supernova evolution. As an example, we focus
on the detection of $\bar\nu_e$ through capture on protons,
$\bar\nu_e+p\to n+e^+$, which has a threshold of 
$E_{\rm th} \approx 1.8$~MeV. The net emission rate  
$R^>_{\bar\nu_\alpha}$ and average energy 
$\bar E^>_{\bar\nu_\alpha}$ for $\bar\nu_\alpha$
with energy above $E_{\rm th}$ are given in Table~\ref{tab:nu}
for each thermal emission process at the four selected stages of 
stellar evolution. For consideration of detection, we define
an effective energy $\bar E^{\rm det}_{\bar\nu_\alpha}$ through
\begin{align} 
\sigma_{\bar\nu_ep}(\bar E^{\rm det}_{\bar\nu_\alpha}) \equiv 
\frac{\int_{E_{\rm th}}^\infty \sigma_{\bar\nu_ep}(E_\nu)
F_{\bar\nu_\alpha}(E_\nu)dE_\nu}{\int_{E_{\rm th}}^\infty 
F_{\bar\nu_\alpha}(E_\nu)dE_\nu}=\frac{1}{R_{\bar\nu_\alpha}^>}
\int_{E_{\rm th}}^\infty \sigma_{\bar\nu_ep}(E_\nu)
F_{\bar\nu_\alpha}(E_\nu)dE_\nu, 
\label{eq:Edet}
\end{align}
where $\sigma_{\bar\nu_ep}(E_\nu)\propto(E_\nu-\Delta)
\sqrt{(E_\nu-\Delta)^2-m_e^2}$\,, with $\Delta=1.293$~MeV
being the neutron-proton mass difference, is the cross section 
for capture of $\bar\nu_e$ with energy $E_\nu$. 
Note that $\bar E^{\rm det}_{\bar\nu_\alpha}$ is also introduced for 
$\bar\nu_x$ in consideration of flavor oscillations between
$\bar\nu_x$ and $\bar\nu_e$. If flavor oscillations are independent
of energy, then the net emission rate $R^>_{\bar\nu_\alpha}$
above the detection threshold contributes to the 
$\bar\nu_e+p\to n+e^+$ event rate in proportion to the product of
$R^>_{\bar\nu_\alpha}$ and the detection cross section
at a single energy $\bar E^{\rm det}_{\bar\nu_\alpha}$.
This provides an efficient way to estimate the event rate
without referring to the detailed emission spectra.
The value of $\bar E^{\rm det}_{\bar\nu_\alpha}$ for each
thermal emission process is also given in Table~\ref{tab:nu}.
It can be seen that $\bar E^{\rm det}_{\bar\nu_\alpha}$ for all 
thermal emission processes increase somewhat as the star ages
and that $\bar E^{\rm det}_{\bar\nu_\alpha}$ for pair annihilation
and photo-neutrino emission increase significantly immediately
prior to core collapse. These increases favor the detection of
$\bar\nu_e$ from later stages of stellar evolution because 
$\sigma_{\bar\nu_ep}(\bar E^{\rm det}_{\bar\nu_\alpha})$
increases sharply for $\bar E^{\rm det}_{\bar\nu_\alpha}$ close
to $E_{\rm th}$. Note that $\bar E^{\rm det}_{\bar\nu_e}\approx
\bar E^{\rm det}_{\bar\nu_x}$ except for the case of pair 
annihilation immediately prior to core collapse.

We have given the relevant information for both $\bar\nu_e$
and $\bar\nu_x$ in Table~\ref{tab:nu} in order to estimate the effect
of $\bar\nu_e\rightleftharpoons\bar\nu_x$ flavor transformation
caused by the MSW mechanism in massive stars. 
When flavor evolution is adiabatic, the survival probability $p$ of 
$\bar\nu_e$ is insensitive to neutrino energy and can be estimated as 
$p_{\rm NH} = \cos^2\theta_{12} \cos^2\theta_{13} \approx 0.7$ for the 
normal mass hierarchy (NH) and 
$p_{\rm IH} = \sin^2\theta_{13} \approx0.025$ 
for the inverted mass hierarchy (IH) \cite{matter}, where
$\theta_{12}$ and $\theta_{13}$ are the vacuum mixing angles.
The $\bar\nu_e$ event rate is proportional to
$p R_{\bar\nu_e}^>\sigma(\bar E_{\bar\nu_e}^{\rm det}) 
+ (1-p) R_{\bar\nu_x}^>\sigma(\bar E_{\bar\nu_{x}}^{\rm det})$.
Table~\ref{tab:nu} shows that $R_{\bar\nu_\alpha}^>$ is comparable 
for all thermal emission processes only for the stage immediately 
prior to core collapse, and $R_{\bar\nu_\alpha}^>$ for pair annihilation 
is always the largest for all the previous stages. Taking into account
that pair annihilation also has the highest 
$\bar E^{\rm det}_{\bar\nu_\alpha}$ (see Table~\ref{tab:nu}), 
we conclude that it is the 
dominant source for the $\bar\nu_e$ signal from the thermal emission 
processes. 

\section{Comparison of energy loss rates for thermal neutrino emission processes}
\label{sec:loss}
\begin{figure*}
\centering
\includegraphics[width=0.45\textwidth]{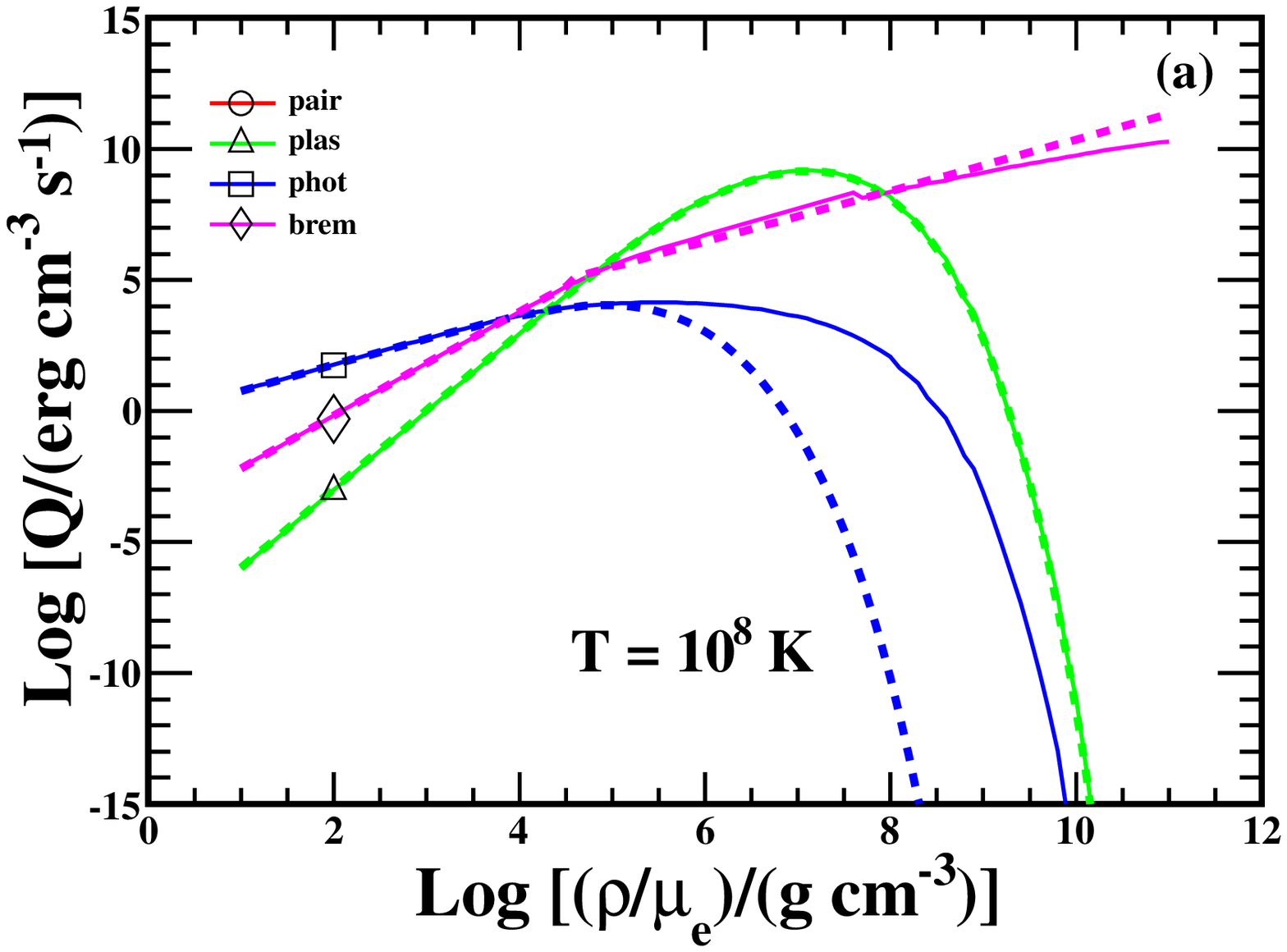} 
\includegraphics[width=0.45\textwidth]{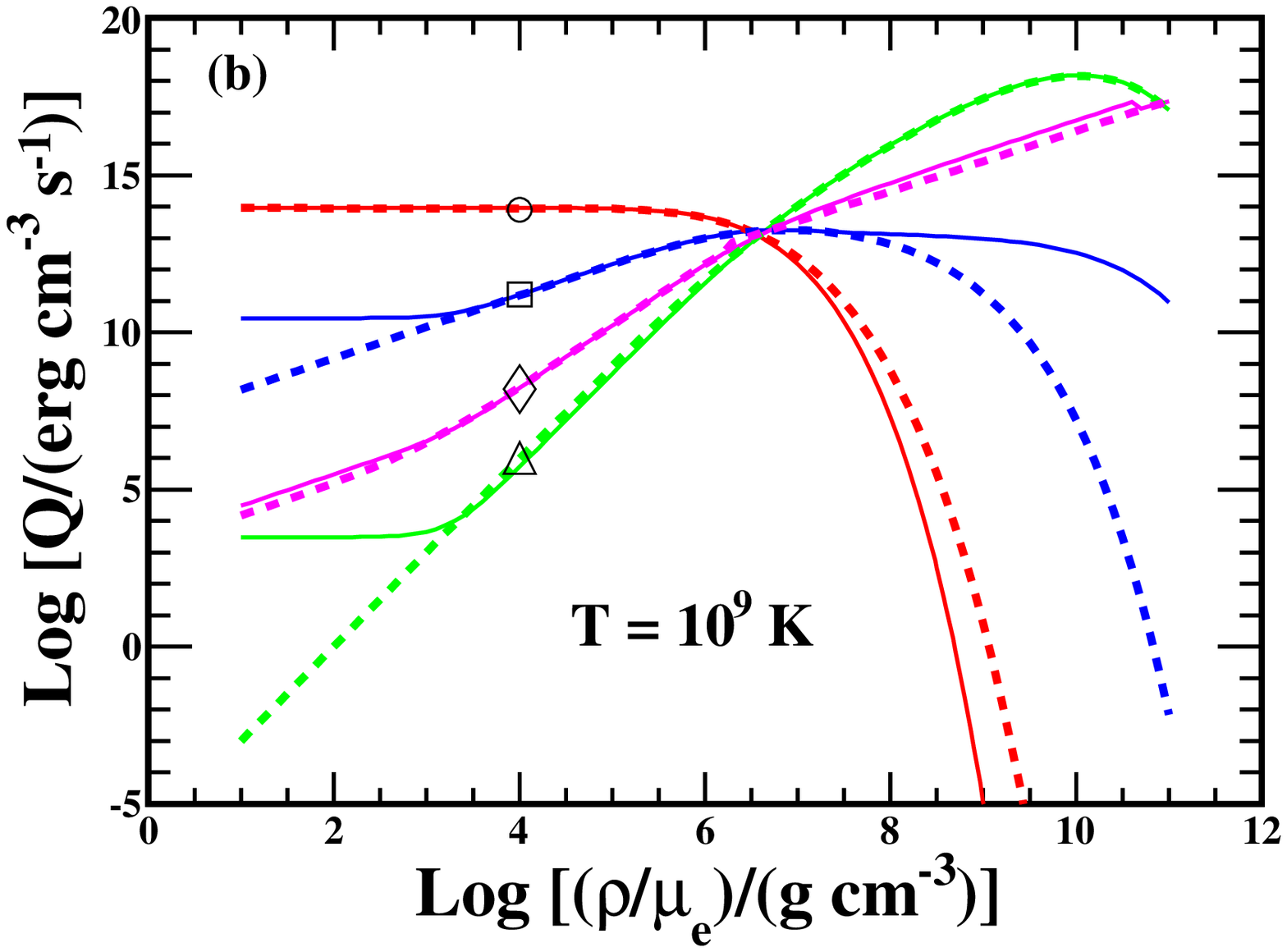}  \\
\includegraphics[width=0.45\textwidth]{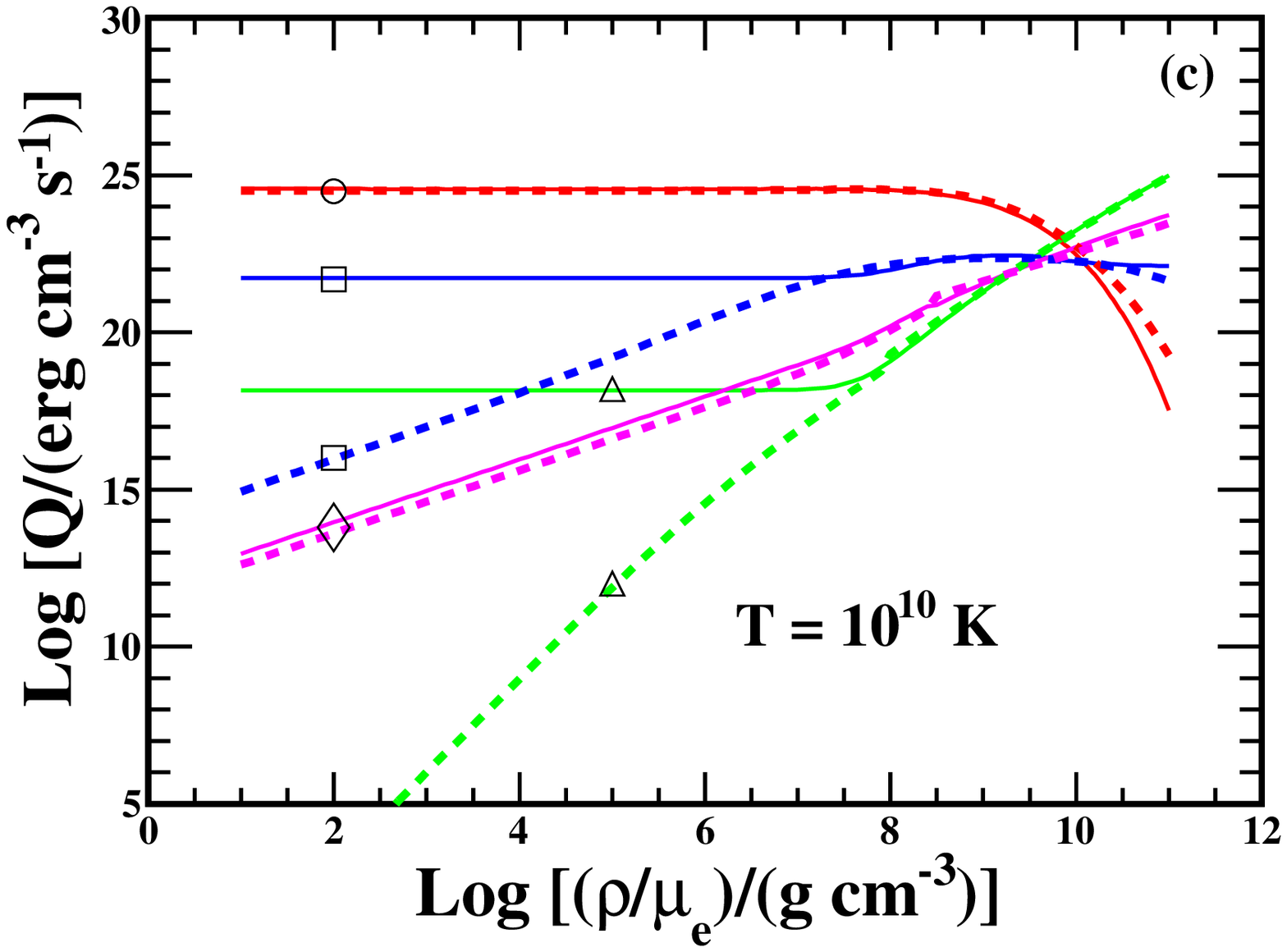} 
\includegraphics[width=0.45\textwidth]{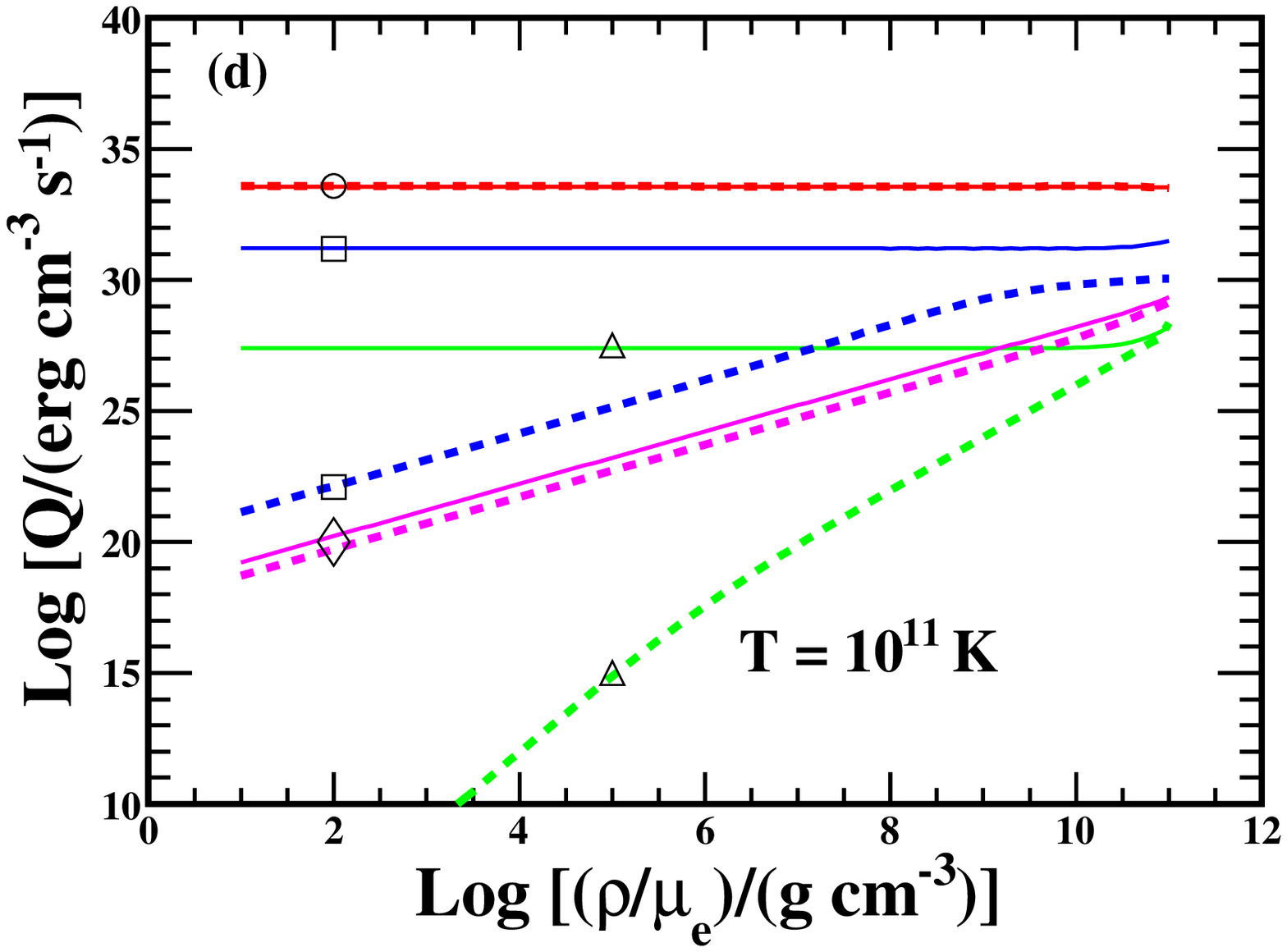} 
\caption{Comparison of energy loss rates $Q$ as functions 
of $\rho/\mu_e$ at (a) $T=10^8$, (b) $10^9$, (c) $10^{10}$,
and (d) $10^{11}$~K for $e^\pm$ pair annihilation (``pair'', $\bigcirc$), 
plasmon decay (``plas'', $\triangle$), photo-neutrino emission (``phot'', $\square$), 
and bremsstrahlung neutrino emission (``brem'', $\lozenge$).
Solid curves are our calculated results while dashed curves
are from the fitting formulas of Ref.~\cite{Itoh96a}. 
Note that $e^\pm$ pair annihilation is highly suppressed 
at $T = 10^8$ K and the corresponding energy loss rate 
is not shown in (a).}
\label{fig:Qtot} 
\end{figure*}

It is straightforward to calculate the total neutrino energy loss rate
per unit volume
\begin{align}
Q = \sum_{\alpha=e, \mu, \tau} 
\int E_\nu[F_{\nu_\alpha}(E_\nu) + F_{\bar\nu_\alpha}(E_\nu)]dE_\nu 
\end{align}
for bremsstrahlung and other thermal emission processes.
For simplicity, we assume an OCP composed of $^{56}$Fe for
calculating the rates for bremsstrahlung neutrino emission in this section.
As the effects of ionic correlations are not very sensitive to composition
(see Fig.~\ref{fig:spec_ionic}), results for a different composition can be 
approximately obtained from those for $^{56}$Fe through scaling with 
$Z^2/A$ [see Eq.~(\ref{eq:brem_spec})].
In Fig.~\ref{fig:Qtot}, we compare our calculated total energy loss rates for 
individual processes with the fitting formulas \cite{Itoh96a} widely used in
stellar evolution models. We show $Q$ as a function of $\rho/\mu_e$
(between 10 and $10^{11}$~g~cm$^{-3}$) for $T=10^8$, $10^9$, 
$10^{10}$, and $10^{11}$~K, respectively. It can be seen that our
results for bremsstrahlung neutrino emission are in good agreement
with the fitting formulas, which provides an indirect check on the 
soundness of our energy-differential rates. The small differences
come from the following several factors. We have used a more recent 
and slightly different structure factor $S_\Gamma(|{\bf k}|)$ and a more
general static dielectric function $\epsilon(|{\bf k}|)$. We have also
included the contributions from positron-nucleus bremsstrahlung.
In addition, the fitting formulas of Ref.~\cite{Itoh96a} have intrinsic 
uncertainties in reproducing the underlying numerical results.        

In consideration of the total neutrino energy loss rates for 
pair annihilation, plasmon decay, and photo-neutrino emission, 
we note that by design, the fitting formulas for a process are 
generally only accurate in the region where this process 
dominates. This accounts for the large discrepancies between 
our results and the fitting formulas in the regions where the latter
fail, especially for plasmon decay and photo-neutrino emission. 
However, when summed over all thermal emission processes,
our results are consistent with the fitting formulas within 5\%--10\%.
Our results are in good agreement with the more up-to-date studies \cite{Dutta03,Ratkovi03,Misiaszek05,Odrzywolek07,Kato15}. 
These and our calculations have used improved treatment of plasmon 
dispersion relations and electrostatic screening for the relevant
processes. The corresponding results are more accurate and should 
be used instead of the fitting formulas when individual thermal emission 
processes are of concern.

\begin{figure}
\includegraphics[width=0.5\textwidth]{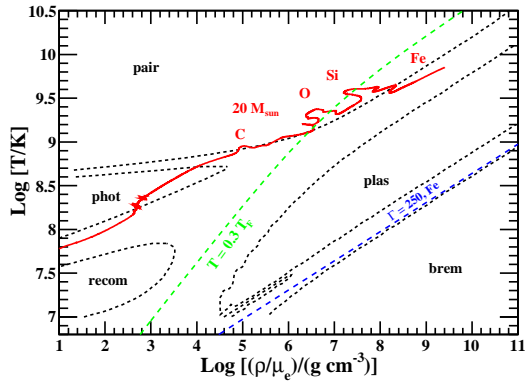}  
\caption{Domain of dominance where a thermal neutrino emission 
process contributes at least 90\% of the total energy loss rate.
Our calculated rates are used for $e^\pm$ pair annihilation (``pair''), 
plasmon decay (``plas''), photo-neutrino emission (``phot''), and 
bremsstrahlung neutrino emission (``brem'') while the fitting formulas 
of Ref.~\cite{Itoh96a} are used for the recombination process (``recom'').
An OCP composed of $^{56}$Fe is assumed for calculating the rate
for bremsstrahlung neutrino emission. Similar to Figs.~\ref{fig:regions}
and \ref{fig:gamma}, also shown are the evolutionary track of the 
central temperature and density for a $20\,M_\odot$ star, 
the curve for $T=0.3T_F$, and the contour for $\Gamma=250$ 
for an OCP composed of $^{56}$Fe.} 
\label{fig:dom}
\end{figure}

We define the domain of dominance for a process as the region in 
the $(T,\rho/\mu_e)$ space where this process contributes at least
90\% of the total neutrino energy loss rate summed over all the 
thermal emission processes. These domains are shown in
Fig.~\ref{fig:dom} based on our results except for the recombination 
process, for which the fitting formulas \cite{Itoh96a} are used.
The energy loss rate for pair annihilation is very sensitive to 
temperature and density. It dominates when the temperature is 
sufficiently high for producing $e^\pm$ pairs and the density is
sufficiently low that positrons are not suppressed by degeneracy.
When electrons are strongly degenerate, plasmon decay, 
photo-neutrino emission, and especially pair annihilation are 
suppressed. In this case, bremsstrahlung neutrino emission 
becomes dominant. When this occurs, ionic correlations are
important and can reduce the energy loss rate by a factor of 
$\sim 2$--10. We note that plasmon
decay dominates in two regions. The transverse decay modes
play a key role in the larger region while the longitudinal decay
mode takes over in the much smaller region.

\section{Discussion and Conclusions}     
\label{sec:disc}
We have presented a detailed derivation of the energy-differential rate 
for neutrino emission from electron-nucleus bremsstrahlung (Sec.~\ref{sec:brem}),
taking into account the effects of electron screening and ionic correlations. 
We have compared the energy-differential and the net rates, as well as the 
average $\bar\nu_e$ and $\bar\nu_x$ energies, for this and 
other thermal neutrino emission processes over a wide range of temperature 
and density (Sec.~\ref{sec:spec}). We have also compared our updated 
energy loss rates for individual thermal neutrino emission processes with the 
fitting formulas of Ref.~\cite{Itoh96a} and determined the temperature and 
density domain in which each process dominates (Sec.~\ref{sec:loss}). We find that similar to plasmon decay and photo-neutrino emission, bremsstrahlung mostly produces sub-MeV neutrinos during the pre-supernova evolution of massive stars. Our results on the neutrino energy loss rates are in good agreement with previous studies.

As discussed by previous studies 
\cite{Odrzywolek04,Odrzywolek10,Kato15,KamLAND15,JUNO,Yoshida:2016}, 
neutrino emission during the
pre-supernova evolution of massive stars can provide a potential test of 
stellar models or at least give advance warning for core-collapse supernovae.
While neutrino emission from $\beta^\pm$ decay and $e^\pm$ capture 
\cite{Patton15,Odrzywolek09} should be taken into account for a full study, 
we expect that $\bar\nu_e$ signals from the thermal processes discussed here 
always dominate except for the last hour or so prior to a supernova explosion. 
Figure~\ref{fig:dom} and Table~\ref{tab:nu} serve as approximate guides to 
the relative importance of each thermal neutrino emission process during 
the pre-supernova evolution of massive stars. With the largest 
$\bar E_{\bar\nu}^{\rm det}$ and $R^>_{\bar\nu}$, pair annihilation 
is always the dominant source of pre-supernova $\bar\nu_e$ signals 
for massive stars during core C burning and afterwards. For bremsstrahlung 
neutrino emission, we note that its domain of dominance is far from the 
evolutionary track of the central temperature and density for a $20\,M_\odot$ star, 
and therefore expect that it contributes only a small fraction of the $\bar\nu_e$ events. 
However, this domain overlaps with the conditions encountered during the cooling of 
neutron stars produced by the core collapse of massive stars. 
We refer readers to Refs.~\cite{Kaminker98,Yakovlev00} for more detailed 
discussions of bremsstrahlung neutrino emission relevant for neutron star cooling.

In general, significant thermal neutrino emission occurs throughout the hot and 
dense interior of a massive star during core C burning and afterwards. A proper 
estimate of the $\bar\nu_e$ signal from the pre-supernova evolution of the star 
requires a model that gives the radial profiles of temperature, density, and 
composition as well as the corresponding time evolution. We refer readers to 
Ref.~\cite{Kato15} for a detailed study on the $\bar\nu_e$ signals from pair 
annihilation and plasmon decay using models for three stars of 8.4, 12, and 
$15\,M_\odot$, respectively. We plan to carry out a systematic study including 
more massive stars and taking into account neutrino oscillations in the near future. 

\acknowledgments
We thank Alexander Heger for providing us with models of massive stars. This work was supported in part by the U.S. DOE Grant No. DE-FG02-87ER40328 and by a key laboratory grant (No. 11DZ2260700) from the Office of Science and Technology in Shanghai Municipal Government. 
      
\appendix
\setcounter{equation}{0}
\section{Quantities in the effective squared matrix element}
\label{sec:app_sqamp}
The quantities $I_i^B\ (i=1,2,3)$ in Eq.~(\ref{eq-m2eff})
for the effective squared matrix element $|{\cal M}|_{\rm eff}^2$ are 
defined as 
\begin{align} 
I_1^B
=&-\frac{2}{\beta_1}(c_1 k^2 + 2c_2c_5 - 4c_3 c_{12})
+\frac{2}{\beta_2}[4(c_3- 4c_5)c_9-c_1k^2-2c_2c_5] 
-\frac{k^2 + 4c_{10}}{\beta_1^2}(c_1c_3+8c_6^2)  \nonumber \\
& -\frac{k^2+4c_7}{\beta_2^2}[c_1c_3 + 8c_6^2+4c_5(2c_5-c_3)] 
+\frac{2}{\beta_1\beta_2}\{4c_2c_5^2 + 4c_{11} [c_3(2c_5-c_1)-8c_6^2] 
\nonumber \\
&+ k^2[c_1^2  + 2c_3c_5 + 
c_1(k^2-4m_e^2-4c_6 + 2c_7 +4c_8 + 4c_9 + 2c_{10} - 4c_{12})
\nonumber \\ 
& - 8c_6^2 -8m_e^2(c_8+c_9-c_{12})]\} 
+c_1\left(\frac{\beta_2}{\beta_1} + \frac{\beta_1}{\beta_2}\right) -8c_8, 
\label{eq:IB_1}
\end{align}

\begin{align}
I_2^B=&-2m_e^2\left\{4c_5\left(\frac{1}{\beta_1}+\frac{1}{\beta_2}\right)
+\frac{c_1}{\beta_1^2}(k^2+4c_{10}) + \frac{c_1}{\beta_2^2}(k^2 +4c_7)
\right.\nonumber \\ 
&\left.- \frac{2}{\beta_1\beta_2}[ k^2 (c_1+4c_8+4c_9 - 4c_{12}) 
+ 4c_5^2 - 4c_1c_{11}] \right\},
\label{eq:IB_2}
\end{align}

\begin{align}
I_3^B =& -\frac{4c_1(c_4+4c_{12})}{\beta_1} + \frac{4c_1}{\beta_2}(c_4-4c_9)
+ \frac{2(k^2+4c_{10}) c_1c_3}{\beta_1^2} + 
\frac{2c_1(k^2+4c_7) (c_3 - 4 c_5)}{\beta_2^2} \nonumber \\
&+\frac{4c_1}{\beta_1\beta_2} 
\Big[4(c_3-2c_5)c_{11} + k^2 (-c_3 +2c_5 + 2c_7 - 2c_{10}  + 4c_9+4c_{12}) \Big]
+ 2c_1\left(\frac{\beta_2}{\beta_1}-\frac{\beta_1}{\beta_2}\right),
\label{eq:IB_3}
\end{align}
where
\begin{align}
&c_1\equiv P^2\equiv(p+k-p')^2,\;\;\; 
c_2\equiv P^2-2m_e^2,\;\;\;
c_3\equiv P^2+4p^\prime \cdot q, 
 \nonumber \\
&c_4\equiv k^2+2k\cdot q,\;\;\;
c_5\equiv k\cdot q,\;\;\;
c_6\equiv p^\prime \cdot q, \nonumber \\
&c_7 \equiv (p^\prime \cdot \epsilon_B)^2,\;\;
c_8\equiv (q \cdot \epsilon_B)^2,\;\;
c_9\equiv (p^\prime \cdot \epsilon_B)(q \cdot \epsilon_B), \nonumber \\
&c_{10} \equiv  (p \cdot \epsilon_B)^2,\;\;
c_{11} \equiv  (p\cdot \epsilon_B)(p^\prime \cdot \epsilon_B),\;\; 
c_{12} \equiv  (p\cdot \epsilon_B)(q \cdot \epsilon_B).
\label{eq:brem_amp5}
\end{align}

%

\end{document}